\documentclass{webofc}

\usepackage[varg]{txfonts}   
\usepackage{hyperref}
\usepackage{url}
\hypersetup{colorlinks=true,citecolor=blue,urlcolor=blue,linkcolor=blue}
\begin{document}
\title{A comprehensive view of nuclear shapes, rotations and vibrations from fully quantum mechanical perspectives} 
%
%

\author{\firstname{Takaharu} \lastname{Otsuka}\inst{1,2,3,4}\fnsep\thanks{\email{otsuka@phys.s.u-tokyo.ac.jp}} 
}

\institute{Department of Physics, University of Tokyo, Hongo, Bunkyo-ku, Tokyo 113-0033, Japan 
\and
           RIKEN Nishina Center for Accelerator-Based Science, Hirosawa 2-1, Wako-shi, Saitama 351-0198, Japan
\and
           Institut f\"ur Kernphysik, Technische Universit\"at Darmstadt, D-64289 Darmstadt, Germany
\and
           Grand Acc\'el\'erateur National d’Ions Lourds, CEA/DRF-CNRS/IN2P3, Bvd Henri Becquerel, F-14076 Caen, France
                                }

\abstract{The nuclear quadrupole collective states at low excitation energies are described in a novel, fully quantum mechanical and systematic manner as compared to traditional pictures initiated by Aage Bohr.  
The ellipsoidal shapes are shown to be triaxial in virtually all strongly deformed nuclei, in contrast to the Ansatz of axially symmetric shapes.  The rotational bands of such triaxially deformed nuclei are described in a fully quantum mechanical way, i. e., without resorting to quantized free rotation of rigid body.  The excitation energies within a rotational band, exhibiting the $J(J+1)$ dependence on angular momentum $J$, are shown to basically represent the change of binding energies due to nuclear forces.  This differs from the interpretation \'a la Aage Bohr as rotational kinetic energies.
The $K$ quantum numbers are shown to be practically conserved for triaxial ellipsoids, which turned out to be a real but positive surprise to many people in the field.  The so-called $\gamma$ bands are shown to be $K$=2$^+$ rotations rather than $\gamma$-vibrations, leading to a nice description of the so-called $\gamma\gamma$ 4$^+$ state as a $K$=4$^+$ rotation.  Vibrational modes are also shown to emerge in this study.  Thus, the whole picture of low-energy quadrupole collective motion of heavy nuclei has been renewed in a fully quantum mechanical fashion, which differs from the traditional picture but appears to be simpler and more natural.  
}
\maketitle
\section{Introduction}
\label{intro}


The nuclear shape is deformed to various ellipsoids in many nuclei, in particular, heavy ones.  
Such deformation was discussed first by J. Rainwater  \cite{rainwater1950} and A.  Bohr and B. Mottelson \cite{bohr1952,bohr_1953,bohr_mottelson1953,aage_bohr_nobel}.  Properties associated with strong deformation were discussed in many works afterwards as described in textbooks, e.g.  Refs.~\cite{bohr_mottelson_book2,rowe_book,deshalit_book,preston_bhaduri_book,ring_schuck_book,eisenberg_greiner_book1,casten_book}.      

I present, in this paper, the novel outcome of our recent works on such collective properties \cite{epja}, so that those who cannot go through \cite{epja} can still grasp the essence of agendas, ideas and consequences.  As a matter of fact, the nuclear shape and its motion are one of the most important and fundamental subjects to be clarified by nuclear physics.   I will focus on underlying physical properties rather than individual numerical results.  I just note that actual calculations were obtained from the state-of-the-art quantum many-body simulations by the Monte Carlo Shell Model (MCSM) \cite{mcsm2001,shimizu2012}.   To be more in detail, its most advanced version, Quasiparticle Vacua Shell Model (QVSM) \cite{shimizu_2021} was used.    

In the nuclei to be discussed, the number of protons and that of neutrons are assumed to be even numbers.  
All states are of positive parity, but this may not be explicitly mentioned for brevity.

\section{Shapes}
\label{sec:shape}

\subsection{Axial Symmetry and Triaxiality}
\begin{figure}[h]
\centering
\includegraphics[width=12cm,clip]{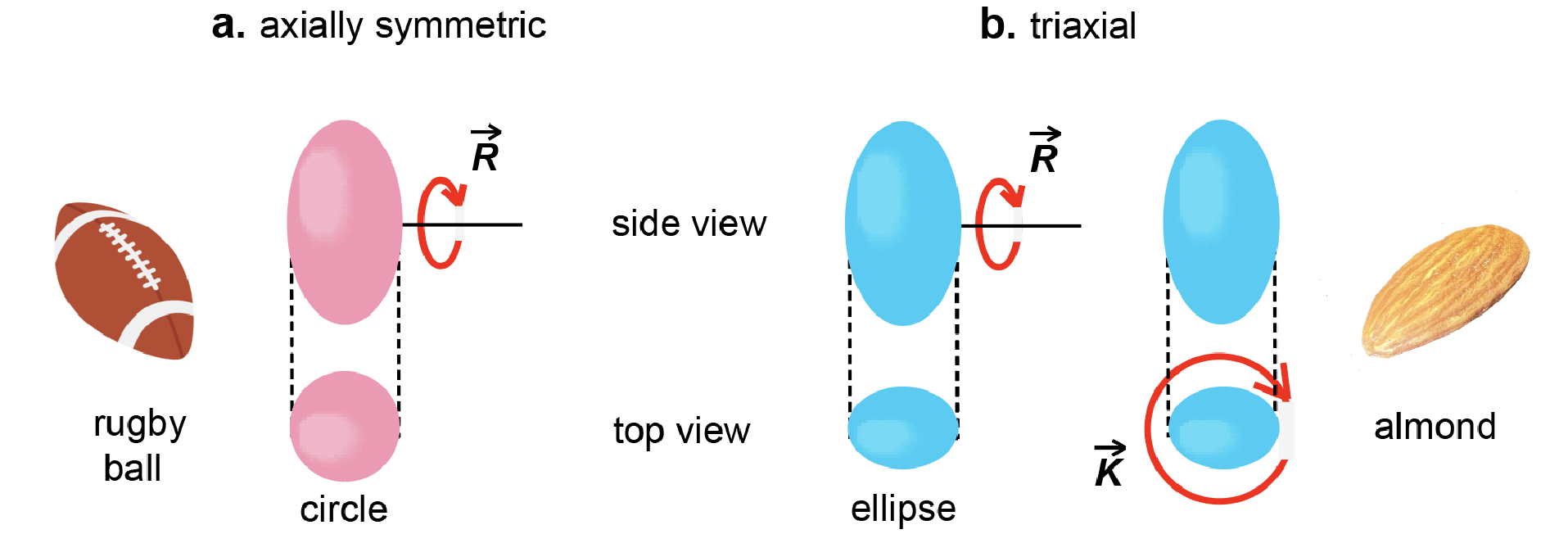}
\caption{Schematic illustrations of the rotations of atomic nuclei in ellipsoidal shapes.
    {\bf a.} axially symmetric and {\bf b.} triaxial nuclear shapes, with associated rotations $\vec{R}$ and $\vec{K}$.  
    The upper part is side views, while the lower part top views.  Intuitive images, rugby ball and almond, are shown
    in both sides.  }
\label{Fig1}       
\end{figure}

We first introduce the concepts of axial symmetry and triaxiality.  
Ellipsoidal shapes of atomic nuclei are schematically shown in Fig.~\ref{Fig1}{\bf a,b}.  Panel {\bf a} shows, in the "top view", the circular cross section perpendicular to the longest axis of the ellipsoid.  This situation is called "axially symmetric deformation", and can be intuitively characterized by a "rugby ball" (see far left of panel {\bf a}).  
Panel {\bf b} stands for a general case with the cross section being an ellipse as seen in its "top view", and is called "triaxial deformation".  The triaxially deformed ellipsoid can be nicely characterized by an "almond" (see far right of panel {\bf b}).   (The case in panel {\bf a} is also called a prolate shape, while an oblate shape does not appear in this article for brevity,).  

In the case of panel {\bf a}, the rotation occurs like $\vec{R}$ about an axis perpendicular to the longest axis.  The rotation about the longest axis does not exist, because the wave function is invariant with respect to this particular rotation (definition of the circle). In the case of panel {\bf b}, a rotation intuitively represented by $\vec{K}$ can occur, besides the rotation $\vec{R}$. 

\subsection{Historical touch}
The strong quadrupole deformation of the nuclear shape was considered, by Bohr and Mottelson since 1950's  \cite{bohr1952,aage_bohr_nobel,bohr_mottelson_book2}, to be axially symmetric, while Rainwater's spheroidal model is also axially symmetric \cite{rainwater1950}.  This basic picture of Bohr and Mottelson has been maintained by the majority of the community up to date, as seen in many textbooks \cite{rowe_book,deshalit_book,preston_bhaduri_book,ring_schuck_book,eisenberg_greiner_book1,casten_book} as well as in other more recent works, for instance, a global survey 
by M\"oller and his co-workers \cite{moller_1995,moller_2006}.   There were certainly different arguments pointing to triaxial shapes.  The work best known among them may be the one by Davydov and his co-workers \cite{Davydov1958,Davydov1959}, while other works can be found, for instance, in ~\cite{yamazaki_1963,sun_2000,sun_2002,Sharpey-Schafer2019}. The triaxial shapes, however, have never been accepted by the community as the basic and standard picture for strongly deformed heavy nuclei.

\subsection{Triaxiality in the nucleus $^{166}$Er as an example}
\label{166Er}

\begin{figure}[h]
\centering
\includegraphics[width=10.5cm,clip]{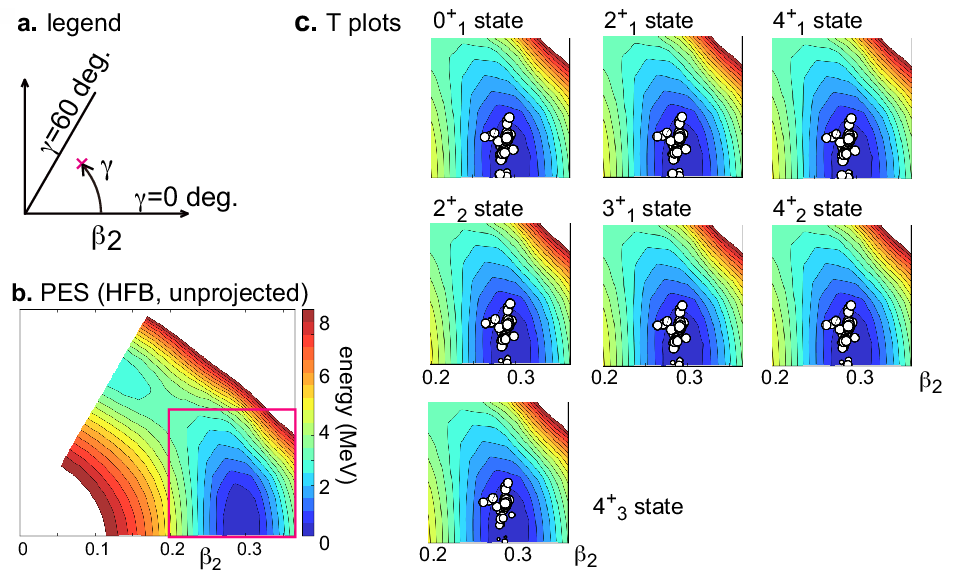}
\caption{{PES and T-plots for $^{166}$Er}.   
    {\bf a} Legend.   
    {\bf b} PES. 
    {\bf c} T-plots of ground and low-lying states for the red-square region of {\bf b}.  
    Based on Fig. 5 of \cite{epja} with kind permission of The European Physical Journal (EPJ).}
\label{Fig5}       
\end{figure}

We looked into the possible problems as to what shapes arise in strongly deformed heavy nuclei, in terms of the Configuration Interaction (CI) calculations with reasonable effective nucleon-nucleon ($NN$) interaction, tested in many other calculations (see references in \cite{epja}).  The level energies and E2 transitions are well reproduced in comparison to experimental data (see Fig. 4 of \cite{epja}).  This salient overall agreement led us to an in-depth analysis of the wave functions.  The MCSM can provide us with the {\bf T-plot} analysis, which visualizes the intrinsic shapes carried by individual basis vectors, by which MCSM eigenstates are expanded \cite{tsunoda_2014}.  
Figure~\ref{Fig5} shows, for the nucleus $^{166}$Er as an example, the Potential Energy Surface (PES) and T-plots with a legend.  Note that this nucleus was taken as an example in the Nobel prize lecture by Aage Bohr 
\cite{aage_bohr_nobel}.  The legend exhibits the deformation parameters $\beta_2$ and $\gamma$, which imply, respectively, the magnitude of quadrupole ellipsoidal deformation and the degree of triaxiality.  The value of $\beta_2$ is about 0.3 for strongly deformed rare-earth nuclei.  Axially symmetric deformation corresponds to $\gamma$=0$^{\circ}$, and $\gamma$=30$^{\circ}$ implies the maximum triaxiality.  

The T-plots in Fig.~\ref{Fig5}{\bf c} show circles corresponding to individual MCSM basis vectors, at their $\beta_2$ and $\gamma$ values.  The overlap probability with the corresponding MCSM eigenstate is indicated by the area of each circle.  The T-plot circles thus exhibit the whereabouts of basis vectors in the $\beta_2$-$\gamma$ plane (see the Fig.~\ref{Fig5}{\bf a}) for each eigenstate.  Thus, the MCSM can give not only the values of physical observables but also deeper insights of the structure of eigenstates.

The T-plot pattern is almost perfectly identical among the 0$^+_1$,  2$^+_1$ and 4$^+_1$ states (i.e., ground band) as a clear indication of the formation of this rotational band.   A similar feature is found among 2$^+_2$,  3$^+_1$ and 4$^+_2$ states (i.e., so-called $\gamma$ band).  
The T-plots slightly differ between the ground band and the $\gamma$ band due to a stretching of the ellipse (see  Fig.~\ref{Fig1}{\bf b}) in going from the ground to $\gamma$ bands. 

The average values of the deformation parameter $\gamma$ obtained from T-plot are 8.2$^{\circ}$, 9.1$^{\circ}$ and 9.5$^{\circ}$ for the ground and $\gamma$ bands and the 4$^+_3$ state (of Fig.~\ref{Fig5}{\bf c}), respectively.  These values are neither small nor results of random fluctuations.  They imply that the energy minimum indeed moves to certain triaxial shapes.
 
\subsection{Triaxiality in other nuclei}
\label{other}

\begin{figure}[h]
\centering
\sidecaption
\includegraphics[width=6.4cm,clip]{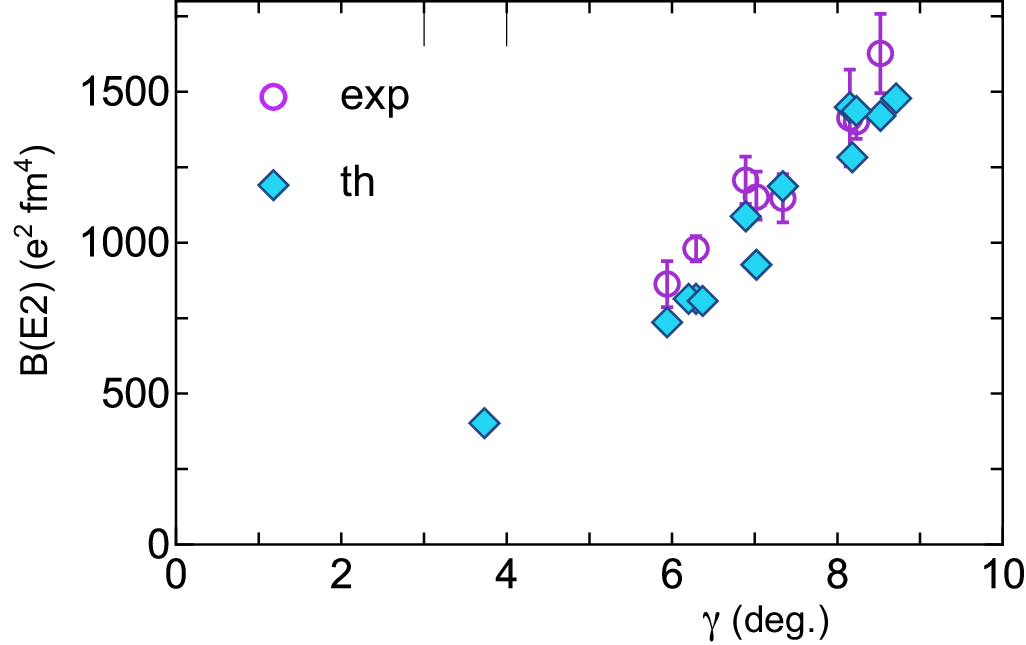}
\caption{Systematics of experimental (open circle) and theoretical (blue diamond) 
    B(E2; 0$^+_1 \rightarrow$ 2$^+_{\gamma}$) values as a function of the deformation 
    parameter $\gamma$.  
The symbols represent, from right to left, $^{164}$Yb ($\gamma$=8.7$^{\circ}$), $^{162}$Er 
    (8.5$^{\circ}$), $^{166}$Er (8.2$^{\circ}$), $^{168}$Hf (8.2$^{\circ}$), $^{164}$Er 
    (8.2$^{\circ}$), $^{164}$Dy (7.3$^{\circ}$), $^{160}$Dy (7.0$^{\circ}$), $^{162}$Dy 
    (6.9$^{\circ}$), $^{166}$Yb (6.4$^{\circ}$), $^{160}$Gd (6.3$^{\circ}$), $^{162}$Gd 
    (6.2$^{\circ}$), $^{158}$Gd (5.9$^{\circ}$), $^{154}$Sm (3.7$^{\circ}$), respectively. 
   The far left blue diamond represents B(E2; 0$^+_1 \rightarrow$ 2$^+_{g\gamma}$) value of 
   $^{154}$Sm. 
   Taken from Fig. 18 of \cite{epja} with kind permission of The European Physical Journal (EPJ).}
\label{Fig18}      
\end{figure}

Substantial triaxial shapes have been identified in the MCSM (or QVSM to be more precise) calculations for 13 heavy rare-earth nuclei as shown in the caption of Fig.~\ref{Fig18} ($^{154}$Sm excluded).   Note that the shapes of these nuclei were traditionally considered to be axially symmetric.  Figure~\ref{Fig18} clearly demonstrates that these nuclei show substantial values of $\gamma$ ($\gtrsim$6$^{\circ}$), and they are correlated remarkably well with the B(E2; 0$^+_1 \rightarrow$ 2$^+_{\gamma}$) value, where 2$^+_{\gamma}$ means the band head of the $\gamma$ band.  Note that the MCSM calculation and experiment agree quite nicely.  We will come back to the case of $^{154}$Sm shortly.

\subsection{Two mechanisms for triaxiality}
\label{mechanism}

\subsubsection{Symmetry restoration with $K$ quantum number}

Two major robust mechanisms for triaxiality have been clarified in the present study \cite{epja}.  One is the symmetry restoration, and we start with an explanation on it.   

\begin{figure}[h]
\centering
\includegraphics[width=12.6cm,clip]{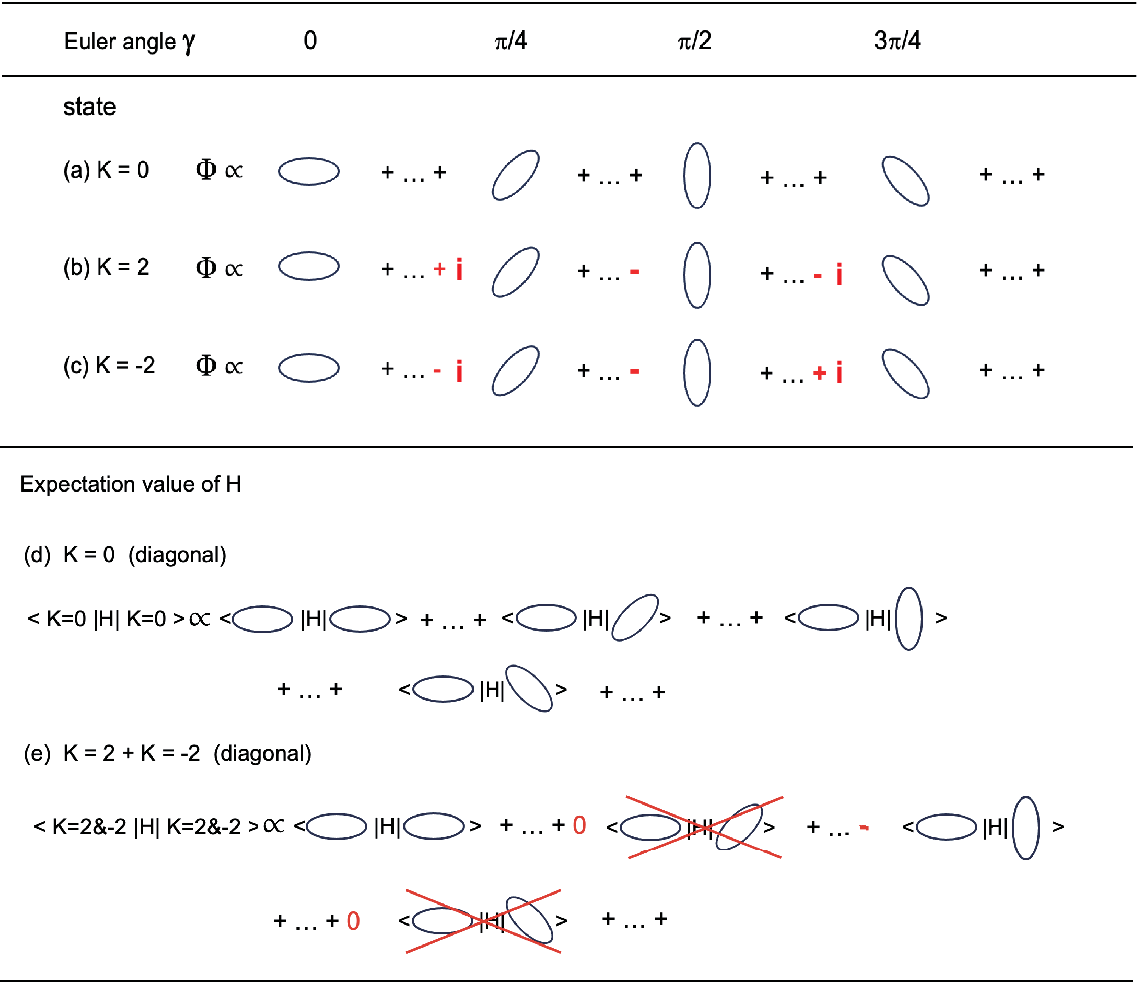}
\caption{Schematic illustrations of K-projected states.
    Taken from Fig. 6 of \cite{epja} with kind permission of The European Physical Journal (EPJ).}
\label{Fig6}     
\end{figure}

The ellipse (i. e., cross section) in Fig.~\ref{Fig1}{\bf b} seen in top view breaks the rotational symmetry in the plane perpendicular to the longest axis.  As the atomic nucleus is an isolated object, the rotational symmetry must be kept.  In order to restore the rotational symmetry in this plane, the ellipse is rotated in the plane and all rotated wave functions are superposed.  Although this superposition is made by an integral with continuously changing angle in the plane, Fig.~\ref{Fig6} shows a discretized visualization.  The amplitudes for the superposition are all equal if the $K$ quantum number shown in Fig.~\ref{Fig1}{\bf b} is $K$=0 (parity is assumed to be positive, also hereafter).  This situation is shown in row (a) of Fig.~\ref{Fig6}.  If such superposition occurs, the diagonal matrix element of the resulting superposed state is given by the summation shown in row (d), where the rotation is discretized similarly.  The first term on the right-hand-side (RHS) is nothing but the mean field value, i.e., no $K$ projection.  The next term with the ket state tilted by $\pi/4$ is considered to have the same sign as the first term for usual effective $NN$ interactions, i.e., an attractive effect.  The same argument holds for other terms, while the magnitude will decrease as the rotation angle increases.  The matrix element of the left-hand-side (LHS) will be lowered by this mechanism.  In other words, the $K$=0 projected state should have a lower energy than the mean-field energy.  This mechanism vanishes if the ellipse is replaced by a circle, as the rotation does not yield different states.  Having more detailed mathematical processes including the norm, the ellipse, or the triaxiality, always results in certain lowering of the energy of $K$=0 state, or increases its binding energy, for reasonable $NN$ interactions.  Thus, as far as ground and low-lying states are concerned, the present symmetry restoration is a robust mechanism favoring triaxiality, and makes virtually all deformed nuclei triaxial, to modest or large extent.

The situation somewhat differs for $K$=2.  Rows (b) and (c) represent K=2 and K=-2 cases, respectively.  The amplitudes for the superposition varies as the rotation angle in the plane changes.  The matrix element of the Hamiltonian can be considered for the $|K$=2$\rangle$ + $|K$=-2$\rangle$ combination because of the reflection (or time reversal) symmetry, and the result is shown in row (e).   The second term on the RHS then vanishes, and the third term is multiplied by a phase factor of $-1$.  Thus, the terms, which increase the binding energy of $K$=0 state, vanish (e.g., the second term in row (e)) or work against the binding energy (e.g., the third term in row (e)) for $K$=2.  The $K$=2 state certainly gains some binding energies from the angles between 0 and $\pi/4$, but the total gain is less than that of the $K$=0 state.  The splitting due to $K$ thus arises.  This mechanism is related to the practical conservation of $K$ discussed later.

\subsubsection{Tensor force and Hexadecupole central force}

The second mechanism originates in specific components of $NN$ interaction.  Two of such components are of particular importance and relevance.  One is the monopole part of the proton-neutron tensor force.  This lowers triaxial states through the self-organization mechanism involving high $j$ orbitals \cite{otsuka_2019}.  The other crucial component is the hexadecupole part of the proton-neutron central interaction.  Their characteristic effects are demonstrated in \cite{epja} also by utilizing constraint HFB calculations with the same shell-model Hamiltonian.

These specific components contribute to different extents for different nuclei, and substantially enlarge the triaxiality for some nuclei, including those mentioned in the caption of Fig.~\ref{Fig18} except for $^{154}$Sm.   

The symmetry restoration mechanism alone seems to produce definite but modest triaxiality: the $\gamma$  value remains less than about 5$^{\circ}$ in rare-earth nuclei.  The triaxiality in such cases is referred to as {\it small triaxiality}, which occurs in virtually all deformed nuclei, as evident from its derivation.  The second mechanism due to the specific $NN$ forces drives nuclei towards substantial triaxiality such as $\gamma \sim$10$^{\circ}$ or even more.  This case is called {\it medium triaxiality}.  The nuclei shown in Fig.~\ref{Fig18} except for $^{154}$Sm belong to this category.    It is of interest that the tensor monopole interaction, which produces the shell evolution in neutron-rich exotic nuclei \cite{otsuka_2020}, also plays important roles in the triaxiality.

\subsubsection{$^{154}$Sm case and GDR spectrum}

The $^{154}$Sm nucleus has been believed to be a typical example of the axially symmetry, in conventional picture.  In the present view, $^{154}$Sm is a good showcase where only the symmetry restoration mechanism works for the triaxiality.   Indeed, the T-plot for $^{154}$Sm exhibits $\gamma\sim$3.7$^{\circ}$, which points to {\it small triaxiality}.  This $\gamma$ value is consistent with the $\gamma$ value suggested by a recent experiment on the GDR spectrum \cite{kleemann_2025}, as the first evidence of finite triaxiality in the ground state of $^{154}$Sm.   
 Very interestingly, for $^{154}$Sm, the so-called $\beta$ and $\gamma$ bands belong to the category of {\it medium triaxiality} with $\gamma\sim$13$^{\circ}$ according to their T-plot.  Thus, different triaxialities coexist in $^{154}$Sm at low excitation energies.   
 
\subsection{$\gamma$ band as $K$=2 rotation of triaxial ellipsoid}
\label{gamma band}

The 2$^+_{\gamma}$ state was identified by Aage Bohr as $J$=2$^+$ vibrational one-phonon excitation from the $J$=0$^+$ ground state \cite{aage_bohr_nobel}.  This interpretation has been kept by (the majority of) the community up to date.   The present work shows that the 2$^+_{\gamma}$ state is the $K$=2 rotation of triaxial ellipsoid, and related E2 properties are calculated in good agreement with experiment.  Furthermore, the so-called $\gamma\gamma$ 4$^+$ state is $K$=4 rotation of triaxial ellipsoid, and its T-plot is shown at the bottom of Fig.~\ref{Fig5}{\bf c}.   Recently, vibrational excitations are clarified on top of the present low-lying states, as described in \cite{tsunoda_2025}.  I skip further description of it, because of the length limitation.

\section{Rotation}
\label{sec:rotation}


\subsection{Historical touch}
\label{rotation_history}

\begin{figure}[h]
\centering
\includegraphics[width=12.9cm,clip]{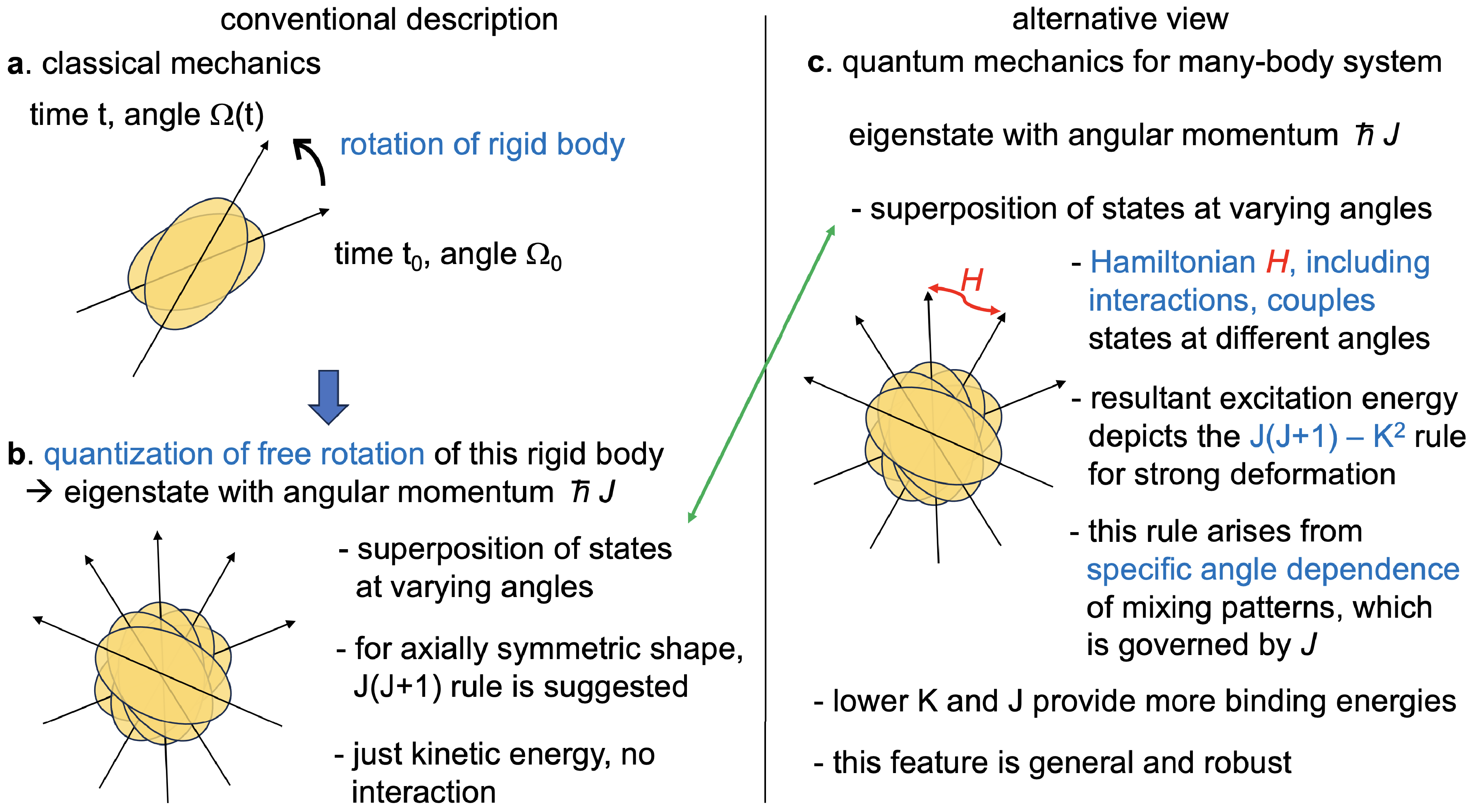}
\caption{Schematic views and descriptions of the rotation.
    {\bf a.} Classical mechanical view.  {\bf b.} Quantization of freely rotating rigid body. 
    {\bf c.} View of the quantum mechanical system composed of many constituents with angular momentum $\hbar J$.  The J(J+1)-K$^2$ rule arises. 
     The green arrow indicates similarity in the wave function, but the energy comes from different 
     origins.     Taken from Fig. 24 of \cite{epja} with kind permission of The European Physical Journal (EPJ).}
\label{Fig24}       
\end{figure}

The rotational motion of a rigid body in the classical mechanics is sketched in Fig.~\ref{Fig24}{\bf a}.
The time evolution of this motion is given by Newtonian equation.  In the quantum mechanics, the angular momentum of the free rotation of a rigid body is quantized as depicted in panel {\bf b}.  The excitation energy is then given by  rotational kinetic energy.   For the angular momentum of the rigid body being $\hbar J$, this rotational kinetic energy is proportional to $(\vec{J}\cdot \vec{J})$, if the rigid body is axially symmetric.   Its eigenvalues are given by $J(J+1)$.     

In the picture of deformed nuclei given by Aage Bohr, the nucleus is described as a deformed object with a fixed shape.  It was then treated as a rigid body.  The above argument can then be utilized.  The so-called Bohr Hamiltonian contains the kinetic term representing the rotational motion of this rigid body about three principal axes (see e.g.,  \cite{ring_schuck_book}).  By assuming the axial symmetry, the $J(J+1)$ rule of excitation energies within a rotational band appears, providing a perfect explanation for the origin of observed level energy regularity ($\propto J(J+1)$) in many nuclei.   This axially-symmetric rigid-body identification for deformed nuclei seems to be one of the major elements of the Nobel prize of physics in 1975, and has remained as a central view of nuclear rotational bands of (the majority of) the community.


\subsection{Fully quantum mechanical description of rotational bands}
\label{rotational bands}

I here present another view which does not need a classical (or semi-classical) picture/interpretation such as rigid-body rotation.
In quantum mechanics, the rotational band can be defined as a set of states projected onto a given angular momentum from a common (relatively simple) intrinsic state.  This idea itself is not new, but we start from it and re-formulate the description of rotational bands, staying inside quantum mechanics. 

Following the angular-momentum projection method with Wigner's D function (see eq.(9) of \cite{epja} where \cite{ring_schuck_book} is cited for it), this can be done.  Sect. 5 of \cite{epja} presents how the $J(J+1)$ dependence of the excitation energies can be obtained only by pursuing the projection procedures.  Actually, for $K$=0 cases, the $J(J+1)$ rule naturally arises as the next-to-leading order (NLO) in the polynomial expansion of the $d$ function in terms of ($\cos\beta$ -1), where $\beta$ denotes an Euler angle in the projection integral.   (The leading order is the 0-th power term, giving the band-head energy.)   For strong deformation, this NLO term is sufficient in the polynomial expansion, as explained in detail in \cite{epja}, including some history.  For general $K$ values, more general $\{ J(J+1) - K^2 \}$ rule has also been derived in \cite{epja}.  

\begin{figure}[h]
\centering
\sidecaption
\includegraphics[width=5.5cm,clip]{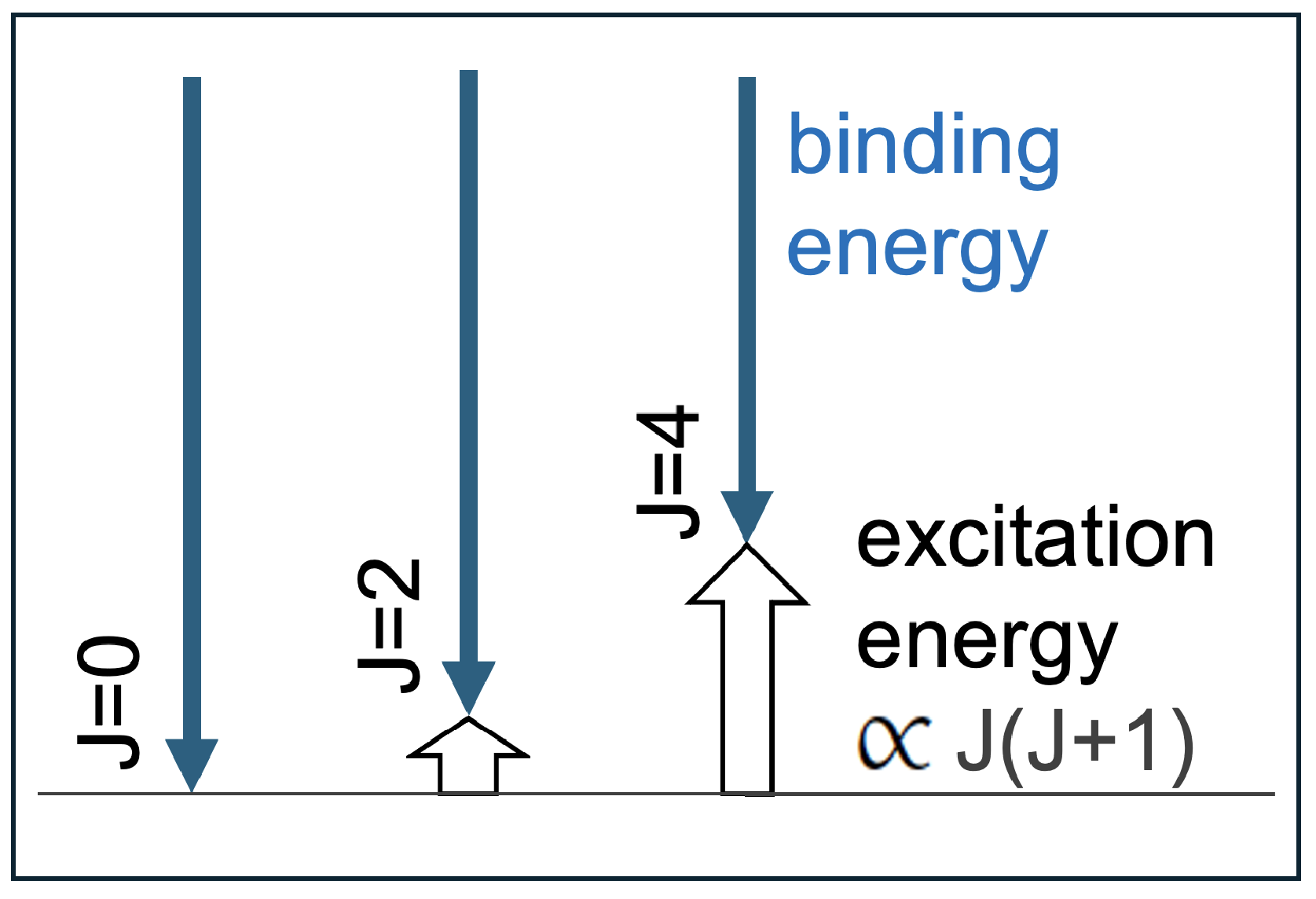}
\caption{A graphical illustration of the origin of the $J(J+1)$ rule of excitation energies within a rotational band.  Blue downward arrows mean binding energies for the states with the angular momentum $J$=0, 2, 4.  The differences from the energy of the $J$=0 state are exhibited by open upward arrows.  The present fully quantum mechanical study indicates their height being proportional to $J(J+1)$, if strong ellipsoidal deformation occurs, irrespectively of triaxiality.  For $K>$0, the $\{ J(J+1)-K^2 \}$ rule is obtained similarly, on top of the $J$=$K$ state.}
\label{J(J+1)}       
\end{figure}

The most important consequence may be the feature displayed in Fig.~\ref{J(J+1)}: the $J(J+1)$ dependence of the excitation energy originates in the $J$-dependent reduction of the binding energy.  This reduction occurs primarily in the contributions of various parts of the nuclear forces, including single-particle energies, two-nucleon forces, three-nucleon forces, {\it etc}.  All parts give the $J(J+1)$ dependence for strong deformation, if one looks at appropriate  differences from the values for the band head ($J$=0 value in Fig.~\ref{J(J+1)} and $J$=$K$ value in general).  The kinetic energy appears to be a tiny fraction, or even can work against (lowering rather than raising).  Note that in Fig.~\ref{J(J+1)}, practical conservation of $K$ quantum number is already incorporated, as it will be demonstrated in subsec~\ref{K}.   

\subsection{$K$ quantum number practically conserved \,-- no $K$ mixing --}
\label{K}

\begin{figure}[h]
\centering
\sidecaption
\hspace{1cm}
\includegraphics[width=5.0cm,clip]{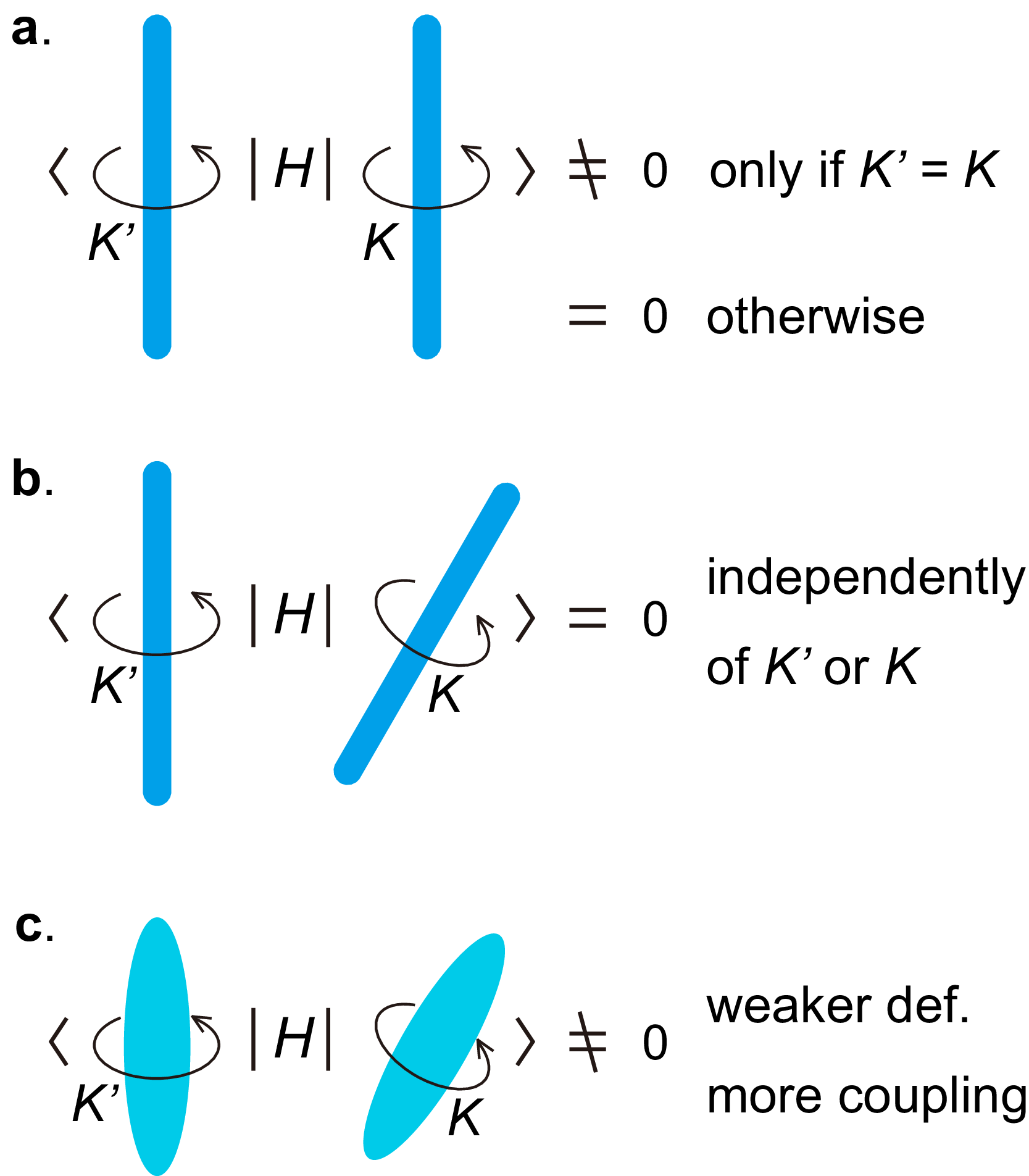}
\caption{{\bf a, b} Schematic illustration of the practical conservation of $K$ quantum number at the limit of strong deformation.  Rods represent strongly deformed triaxial intrinsic states projected onto $K$ or $K'$.  {\bf c} Schematic illustration of the $K$ mixing in weakly deformed nuclei expressed by ellipsoids.  Taken from Fig. 14 of \cite{epja} with kind permission of The European Physical Journal (EPJ).}
\label{Fig14}       
\end{figure}

One of the beliefs by (most of) the community was strong $K$ mixing caused by triaxiality.  I am not aware of the origin of this belief.  Figure~\ref{Fig14}{\bf a, b} depict the basic mechanism practically conserving $K$ quantum number for strongly deformed nuclei: in the projection calculation, matrix elements shown in these panels scale the effects, and can be finite only for $K'$=$K$, i.e., the conservation of $K$.  Contrary to the traditional belief, this feature arises independently of triaxiality, and has been verified numerically with the purity better than 99\% for many cases \cite{epja}.   This feature may be weakened for much weaker deformation as shown in panel {\bf c}.

\section{Concluding remarks}

I would like to mention four points in concluding this proceedings article.

1. Ellipsoidal shapes of atomic nuclei are triaxial.  The degree of the triaxiality varies among nuclei, but all deformed nuclei are triaxial due to the symmetry restoration mechanism, producing {\it small triaxiality} at least.  The tensor force and the hexadecupole central force drive some nuclei to shapes with larger $\gamma$ values, the situation called {\it medium triaxiality}.

2. The excitation energies within a rotational band are dominantly due to the reduction of binding energy produced by nuclear forces.
The $J(J+1)$ dependence appears as a consequence of strongly deformed structure, and the contribution of kinetic rotational energy is minimal or can be even opposite.  The fundamental mechanism drastically differs between the rigid body and the many-body system with interactions among constituents.   The free rotation of a rigid body may not be appropriate for a nucleus as a quantum many-body system, but the transition from quantum to classical pictures of the "rotation" is an open interesting question.

3.  The conventional worry of strong $K$ mixing in triaxial nuclei is irrelevant.  The $K$ quantum number is practically well conserved in triaxial deformed nuclei.  This enables us to always classify rotational bands with $K$, justifying the usual empirical practice.

4.  The 2$^+_{\gamma}$ band is generated by the $K$=2 rotation of triaxial ellipsoid, and there is a similar story for $K$=4.  The resulting observables appear to be consistent with experiment.\\

\noindent
In addition to rotational modes, we developed a new theory for vibrational modes \cite{tsunoda_2025}, but the discussions on it cannot be made here because of the limitation of the length.\\

\noindent
I finally address that the preponderance of triaxiality in heavy nuclei was proposed by a Ukrainian physicist, A. S. Davydov (Crimea 1912 - Kyiv 1993).   
The concept of preponderance of triaxiality turned out to be appropriate from the viewpoint of the present work.  Despite several drawbacks of the Davydov model, I feel that the proposal of this concept should be more appreciated by the community.   Pioneer experimental works with Multiple Coulomb Excitation by D. Cline are to be better appreciated also.\\

\noindent
The author thanks the organizers of this conference for making such an exciting meeting possible.  
He is grateful to Drs Y. Tsunoda, N. Shimizu, Y. Utsuno, T. Abe, H. Ueno and T. Duguet for collaborating on the subjects presented in this talk.  He is grateful to Profs. P. Van Duppen, N. Pietralla, P. Ring and S. Leoni for valuable advices, and to the Humboldt Foundation and the GANIL for supports.  The author also expresses special thanks to Dr. T. Kobori for continuous encouragements.  
This work was supported in part by MEXT as 
``Program for Promoting Researches on the Supercomputer Fugaku'' (Simulation for basic science: from fundamental laws of particles to creation of nuclei, JPMXP1020200105, Simulation for basic science: approaching the new quantum era, JPMXP1020230411), and by JICFuS as well as similar earlier programs.
This work was supported by JSPS KAKENHI Grant Number JP25K00998.

%

\begin{thebibliography}{}
%
%


\bibitem{rainwater1950}
J. Rainwater, 
Nuclear Energy Level Argument for a Spheroidal Nuclear Model,
Phys. Rev. {\bf 79}, 432 (1950).

\bibitem{bohr1952}
A. Bohr, 
The Coupling of Nuclear Surface Oscillations to the Motion of Individual Nucleons, 
Mat. Fys. Medd. Dan. Vid. Selsk. {\bf 26}, 14 (1952).

\bibitem{bohr_1953}
A. Bohr, B. R. Mottelson,
Collective and Individual-Particle Aspects of Nuclear Structure, 
Mat. Fys. Medd. Dan. Vid. Selsk. {\bf 27}, 16 (1953).

\bibitem{aage_bohr_nobel}
A. Bohr, 
Rotational Motion in Nuclei,
{\it Nobel Lectures, Physics 1971--1980}, 
Editor Stig Lundqvist (World Scientific, Singapore, 1992); 
https://www.nobelprize.org/prizes/physics/1975/bohr/facts/.

\bibitem{bohr_mottelson1953}
A. Bohr and B. R. Mottelson, 
Collective and Individual-Particle Aspects of Nuclear Structure, 
Mat. Fys. Medd. Dan. Vid. Selsk. {\bf 27}, 16 (1953).

\bibitem{bohr_mottelson_book2}
A. Bohr and B. R. Mottelson, 
{\it Nuclear Structure} (Benjamin, New York, 1975), Vol. II.

\bibitem{rowe_book}
D. J. Rowe, 
{\it Nuclear collective motion: models and theory} 
(Methuen, London, 1970).

\bibitem{deshalit_book}
A. De Shalit and H. Feshbach, 
{\it Nuclear Structure (theoretical Nuclear Physics)} 
(John Wiley and Sons, New York, 1974).

\bibitem{preston_bhaduri_book}
M. A. Preston and R. K. Bhaduri, {\it Structure of the Nucleus} 
(Addison-Wesley, New York, 1975), Chap. 9.

\bibitem{ring_schuck_book}
P. Ring and P. Schuck, 
{\it The Nuclear Many-Body Problem}
(Springer-Verlag: Berlin, 1980).

\bibitem{eisenberg_greiner_book1}
J. M. Eisenberg and W. Greiner, 
{\it Nuclear Theory}, 3rd ed., 
(North-Holland, Amsterdam, 1987), Vol. I.

\bibitem{casten_book}
R. F. Casten, 
{\it Nuclear structure from a simple perspective}
(Oxford University Press, New York, 2000).

\bibitem{epja}
T. Otsuka, Y. Tsunoda, N. Shimizu, Y. Utsuno, T. Abe, and H. Ueno,
Prevailing triaxial shapes in atomic nuclei and a quantum theory of rotation of composite objects,
Eur. Phys. J. A {\bf 61}, 126 (2025).


\bibitem{mcsm2001}
T. Otsuka, M. Honma, T. Mizusaki, N. Shimizu, and Y. Utsuno,
Monte Carlo shell model for atomic nuclei,
Prog. Part. Nucl. Phys. {\bf 47}, 319 (2001).

\bibitem{shimizu2012}
N. Shimizu {\it et al.}, 
New-generation Monte Carlo shell model for the K computer era, 
Prog. Theor. Exp. Phys. \textbf{2012}, 01A205 (2012).

\bibitem{shimizu_2021}
N. Shimizu, Y. Tsunoda, Y. Utsuno, and T. Otsuka,
Variational approach with the superposition of the symmetry-restored quasiparticle vacua for nuclear shell-model calculations,
Phys. Rev. C {\bf 103}, 014312 (2021).

\bibitem{moller_1995}
P. M\"oller, R. Nix, W. D. Myers, and W. J. Swiatecki, 
At. Data Nucl. Data Tables \textbf{59}, 185 (1995). 

\bibitem{moller_2006}
P. M\"oller, R. Bengtsson, B. G. Carlsson, P. Olivius, and T. Ichikawa, 
Phys. Rev. Lett. \textbf{97}, 162502 (2006). 

\bibitem{Davydov1958}
A. S. Davydov and G. F. Filippov,     
Rotational states in even atomic nuclei, 
Nucl. Phys. \textbf{8}, 237 (1958).

\bibitem{Davydov1959}
A. S. Davydov and V. S. Rostovsky,     
Relative Transition Probabilities between Rotational Levels of Non-axial Nuclei, 
Nucl. Phys. \textbf{12}, 58 (1959).

\bibitem{yamazaki_1963}
T. Yamazaki,
Phenomenological discussion of equilibrium shape of deformed nuclei,
Nucl. Phys. {\bf 49}, 1 (1963).

\bibitem{sun_2000}
Y. Sun, K. Hara, J. A. Sheikh, J. G. Hirsch, V. Vel\'azquez, and M. Guidry,
Multiphonon $\gamma$-vibrational bands and the triaxial projected shell model.
Phys. Rev. C {\bf 61}, 064323 (2000).

\bibitem{sun_2002}
P. Boutachkov, A. Aprahamian, Y. Sun, J. A. Sheikh, and S. Frauendorf, 
In-band and inter-band B(E2) values within the Triaxial Projected Shell Model.
Eur. Phys. J. A {\bf 15}, 455 (2002).

\bibitem{Sharpey-Schafer2019}
J. F. Sharpey-Schafer, R. A. Bark, S. P. Bvumbi, T. R. S. Dinoko, and S. N. T.
Majola, 
``Stiff'' deformed nuclei, configuration dependent pairing and the $\beta$ and $\gamma$ degrees of freedom,
Eur. Phys. J. A {\bf 55}, 15 (2019).

\bibitem{tsunoda_2014}
Y. Tsunoda, T. Otsuka, N. Shimizu, M. Honma, and Y. Utsuno, 
Novel shape evolution in exotic Ni isotopes and configuration-dependent shell structure,   
Phys. Rev. C \textbf{89}, 031301(R) (2014).




\bibitem{otsuka_2019}
T. Otsuka, Y. Tsunoda, T. Abe, N. Shimizu, and P. Van Duppen,   
Underlying Structure of Collective Bands and Self-Organization in Quantum Systems,
Phys. Rev. Lett. \textbf{123}, 222502 (2019).

\bibitem{otsuka_2020}
T. Otsuka, A. Gade, O. Sorlin, T. Suzuki, Y. Utsuno, 
Evolution of shell structure in exotic nuclei, 
Rev. Mod. Phys. {\bf 92}, 015002  (2020).

\bibitem{kleemann_2025}
J. Kleemann, {\it et al.},Gamma decay of the $^{154}$Sm Isovector Giant Dipole Resonance: Smekal-Raman Scattering as a Novel Probe of Nuclear Ground-State Deformation,
accepted by Phys. Rev. Lett. \textbf{134}, 022503 (2025).

\bibitem{tsunoda_2025}
Y. Tsunoda, T. Otsuka, N. Shimizu, T. Duguet, Y. Utsuno and T. Abe, 
Vibrational Modes in Strongly Deformed Nuclei,
arXiv:2507.20275 [nucl-th].


\end{thebibliography}
%
%

\end{document}